\documentclass[preprint]{JHEP}
\usepackage{epsfig}
\usepackage{amssymb,amsfonts}
\relax

\def\be{\begin{equation}}
\def\ee{\end{equation}}
\def\bs{\begin{subequations}}
\def\es{\end{subequations}}

\def\mpl{M_{\rm Pl}}

\def\sp{\;\;\;,\;\;\;}

\newcommand\fverb{\setbox\pippobox=\hbox\bgroup\verb}
\newcommand\fverbdo{\egroup\medskip\noindent%
                        \fbox{\unhbox\pippobox}\ }
\newcommand\fverbit{\egroup\item[\fbox{\unhbox\pippobox}]}
\newbox\pippobox
\def\tr{\tilde\rho}

\def\bbox{\nabla^2}

\def\mt{{\tilde m}}
\def\rct{{\tilde r}_c}
\def \lta {\mathrel{\vcenter
     {\hbox{$<$}\nointerlineskip\hbox{$\sim$}}}}
\def \gta {\mathrel{\vcenter
     {\hbox{$>$}\nointerlineskip\hbox{$\sim$}}}}
\title{Induced Gravity on RS Branes}

\author{ E. Kiritsis, N. Tetradis and T.N. Tomaras\\
Department of Physics, University of Crete, and FO.R.T.H.\\
71003 Heraklion, GREECE\\
{\tt E-mail: kiritsis, tetradis, tomaras@physics.uoc.gr}}

\preprint{\hepth{0202037}}      

\abstract{It is shown that a localized four-dimensional Einstein term,
induced by quantum corrections, modifies significantly the law of gravity in a 
Randall-Sundrum brane world. In particular, the short-distance behavior 
of gravity 
changes from five- to four-dimensional, while,
depending on the values of parameters, 
there can be an intermediate range where 
gravity behaves as in five dimensions.
The spectrum of graviton fluctuations around the brane, 
their relative importance for the gravitational 
force, and the relevance of their emission in the bulk for the brane cosmology 
are analysed. 
Finally, constraints on parameters are derived 
from energy loss in astrophysical 
and particle physics processes.
}

\begin{document}

\maketitle 

\setcounter{equation}{0}

\section{Introduction}

Quantum loop effects due to particles confined on a p-brane will induce 
a p+1-dimensional Einstein term, localized on the p-brane, in addition to 
the usual bulk terms in the effective action. Such a 
``lower-dimensional" 
term, in general 
absent at tree level in string theory and in effective supergravities, 
is certainly allowed by the symmetries of the theory
not broken by the presence of the p-brane. If sizeable, it
can result in major changes in the way the gravitational interaction is 
perceived on the brane.

A phenomenological approach to analyse these effects was followed 
in an example with a 3-brane embedded in a five-dimensional space-time bulk 
and 
the relevant part of the gravitational action parametrized as \cite{dgp,dgkn,dg2}
\begin{equation}
S = M^3\int d^5x\, \sqrt{-g}R+M^3r_c\int d^4 x\sqrt{-\hat g}\hat R\,,
\label{00}
\end{equation}
where ${\hat g}_{\alpha \beta}$, with $\alpha,\beta=0,1,2,3$, is
the induced metric on the 3-brane. When the fifth dimension is assumed
non-compact, the propagator of the graviton, as viewed by an observer on the brane, 
has the form $\sim M^{-3} (p + r_c p^2 /2)^{-1}$, where $p$ is the magnitude of its
four-momentum along the 3-brane. Thus, at length scales $l \sim p^{-1}
\gg r_c$ the gravitational potential behaves as in five dimensions, while in the
opposite limit $l \sim p^{-1} \ll r_c$ it is effectively four-dimensional.

As pointed out in~\cite{dgkn}, an additional difference of this setup from 
standard compactification is that here the analogues of Kaluza-Klein states 
are weakly coupled to the brane fields and experiment provides 
less stringent constraints.
In particular, upon
compactification of the fifth dimension on a circle of radius $R$
with $R\ll r_c$, the gravitational interaction is 
four-dimensional at all scales \footnote{This may also happen without
compactification, as argued in~\cite{Deffayet:2001uk}.}. This can be understood from
the fact that gravity is four-dimensional at length scales
smaller than $r_c$, while the compact circle makes space-time
effectively four-dimensional at length scales larger than $R$ as
well. There is an infinite discrete spectrum of graviton modes, 
with couplings to the brane fields that are
suppressed compared to the usual compactification scenario. This
modifies the appropriate experimental limits coming from solar
system motions and energy loss in stars and supernovae,
and allows values as exotic as $R\simeq 10^{-4} r_c \simeq
10^{16}\,\mathrm{m}$~\cite{dgkn}.

These results are valid for infinitely thin branes. The case of thick branes 
was studied in \cite{ktt}. With $\epsilon$ being the brane thickness and $M$ the 
five-dimensional Planck mass, it was shown 
(a) that at large distances the above picture remains the same, while 
(b) at distances smaller than $~\sqrt{\epsilon/M}$ the Equivalence 
Principle is generically violated. To preserve it, one should fine tune 
the profiles of all particles on the brane. Similar conclusions were presented 
also in \cite{Dubovsky:2001pe}.
Moreover, in \cite{ktt} the effects of $R^2$ terms, both in the bulk and on the brane, were analysed and shown to genericaly modify the behavior 
of gravity at all distances.

Gravity is expected to be induced on D-branes \cite{Iglesias:2001iz}.
Higher-derivative gravitational terms on N=4 D-branes have been calculated
\cite{bbg}.
The induced one-loop Einstein term on the brane was recently  
calculated in the context of string theory \cite{ktt}. 
In particular, it was shown 
that on a collection of D5-branes, with as much as N=2 supersymmetry, an Einstein term
is induced, with a coefficient that can be varied at will by varying 
the compactification moduli.
Moreover, the transition to four-dimensional gravity can happen well below the string scale.
This concrete realization has a much richer set of threshold scales
than the naive five-dimensional model. The five-dimensional Planck
scale appears as a ``mirage'' threshold: physics becomes ten-dimensional, and eventually 
ten-dimensional gravity becomes strong earlier.
Also, it has been pointed out \cite{Corley:2001hg}, that on D-branes 
in bosonic string theory the complete absence of supersymmetry is 
responsible for an induced Einstein term appearing already at tree level.
The effects of that term are important at distances of the order of the string scale.
Finally, the case of induced gravity on branes located in bulk space with more than 
one transverse dimensions has been analysed as well \cite{dg2,ktt,Dvali:2001ae}.
Attemps to use this in model building were also made \cite{Kakushadze:2001bd}.

So far, we have reviewed the physics of the $second$ most important 
gravitational terms in the 
IR, namely the Einstein terms in the bulk and on the brane. We implicitly 
assumed
that the cosmological constants of the bulk and of the brane, which are actually the leading
terms in the IR, are both zero. The purpose of this work is to extend the above
phenomenological analysis and study the effect of non-vanishing cosmological constants.
We will mostly focus on a 3-brane world, embedded in a five-dimensional bulk and we 
will fine tune (\`a la RS \cite{rs}) the two cosmological constants so that the 
brane is flat \footnote{For a recent discussion of exact solutions 
in the five-dimensional context with an induced Einstein term see \cite{Kof}.}. Phrased in
an equivalent way, we will be interested in the effects of the induced 
four-dimensional
Einstein term on the RS scenario. The generic case of unrelated cosmological constants
will be treated elsewhere. 

We have organized our paper in six sections, of which this Introduction is the first.
In Section 2 we describe the effective action relevant to our discussion. We add the 
induced four-dimensional Einstein term to the 
action studied in \cite{rs}, or equivalently, we include the two aforementioned 
cosmological terms in (\ref{00}).
We compute the
graviton propagator as well as the gravitational potential due to a point source on the
3-brane and compare the results with the RS picture.
Section 3 is devoted to the analysis of the spectrum of 
gravitational fluctuations around the brane
background. The couplings of the corresponding eigenmodes to the brane matter are 
obtained here and their relative importance to the gravitational potential is analysed.
The effects of a non-vanishing energy-momentum tensor on the brane and the main features 
of the corresponding cosmology are presented in Section 4. The phenomenon of 
brane energy loss through the decay of ordinary matter to Kaluza-Klein gravitons into
the bulk, together with its implications on astrophysical processes and on the cosmological
evolution, are studied in Section 5. Finally, Section 6 contains our conclusions and a 
discussion of the potential relevance of our results.

\section{Induced gravity in the Randall-Sundrum setup}
\setcounter{equation}{0}

We are interested in the effects of an induced four-dimensional Einstein
term on the Randall-Sundrum (RS) brane configuration \cite{rs}. 
We consider a 3-brane in a five-dimensional bulk with a negative cosmological constant.
Quantum loops of the particles confined on the brane will induce a four-dimensional 
Einstein term on the brane, which, as was pointed out in \cite{dgp,dgkn},
can have important effects 
on the effective gravitational interaction on it.

The relevant part of the low energy 
effective action is parametrized as
\be
S=\int d^5x~ \sqrt{-g} \left( -\Lambda + M^3 R \right)
+\int d^4 x\sqrt{-\hat g} \left( -V_b + M^3r_c \hat R \right),
\label{001}
\ee
where ${\hat g}_{\alpha \beta}$, with $\alpha,\beta=0,1,2,3$,
is the induced metric on the 3-brane. 
The fifth dimension is assumed to be non-compact. 
This corresponds to the limit that the
negative-tension brane is taken to infinity \cite{rs}.
We identify 
$(x,z)$ with $(x,-z)$, where $z\equiv x_4$. However, following the conventions
of \cite{rs} we extend the bulk integration over the entire interval
$(-\infty,\infty)$.
The quantity $V_b$ includes the brane tension as well as 
quantum contributions to the 
four-dimensional cosmological constant. The strength of the 
induced Einstein term
is parametrized in terms of the fundamental mass scale $M$ 
and the length scale $r_c$
\cite{dgp}. 

Of the four a priori parameters $M,\Lambda,V_b$ and $r_c$ of (\ref{001}), one 
$(M)$ sets the scale, while for $V_b^2=-12\Lambda M^3$
the corresponding equations of motion admit the standard
RS solution 
\be 
ds^2=e^{2 A(z)} \eta_{\alpha \beta} dx^\alpha dx^\beta + dz^2,
\label{rs1} 
\ee
with $A(z)=-k|z|$ and $k=-\Lambda/V_b$. We choose $1/k$ and $r_c$, both with dimensions 
of length, to be our two remaining parameters. 
We are interested in the graviton propagator in this background, corresponding
to a perturbation in the metric for which $\eta_{\alpha \beta}$ is replaced by 
$\eta_{\alpha\beta}+h_{\alpha\beta}$ in the above expression. For simplicity 
we will ignore the tensor structure of the graviton field 
and solve instead for the massless scalar 
propagator $G(x,z)$, which is expected to capture 
the essential features of the gravitational correlations and satisfies 
\be
M^3\left[{1\over\sqrt{-g}}\partial_\mu\left(\sqrt{-g}g^{\mu\nu}\partial_\nu
\right)
+\delta(z)
{{r_c}\over\sqrt{-{\hat g}}}\partial_\alpha\left(
\sqrt{-{\hat g}}{\hat g}^{\alpha\beta}
\partial_\beta\right)\right]G(x,z)=\delta^{(4)}(x)\delta(z).
\ee
In the background (\ref{rs1}) in particular, it obeys
\be
M^3 \left( e^{-2A}\bbox_{4}-\partial_z^2-4A'\partial_z+r_c \delta(z)\bbox_4 
\right) G(x,z)=
\delta^{(4)}(x)\delta(z),
\label{4}
\ee
where $\bbox_4$ is the flat Laplacian in four dimensions and the prime denotes 
differentiation with respect to $z$.
Going to momentum space for the four-dimensional part we obtain the 
equivalent equation (we work in Euclidean space with $p^0=-ip^5$)
\be
M^3 \left( e^{-2 A}p^2-\partial_z^2-4A'\partial_z+r_c \delta(z)p^2 \right)
G(p,z)=\delta(z),
\label{5}
\ee
where $p^2=p^2_5+p^2_1+p^2_2+p^2_3$.

The symmetry of the problem implies that the propagator is a symmetric 
function of $z$, i.e. $G(p,z)=G(p,-z)$.
The solution for $z>0$ is given by
\be
G(p,z) = B \, w^2 \, K_2(wp/k)
\label{sol1} 
\ee
with $w\equiv e^{kz}$ and $K_2$ the modified Bessel function.
The multiplicative constant $B$ is fixed by the requirement on the 
solution to satisfy the discontinuity condition
\be
{{\partial G(p,z)} \over{\partial z}}\Big|_{z=0+} - 
{{\partial G(p,z)} \over{\partial z}}\Big|_{z=0-} = 
r_c p^2 G(p,0) - {1\over{M^3}}
\label{disc}
\ee
obtained from (\ref{5}) by integrating both sides in the interval
$(-å, +å)$ and taking the limit $å \to 0$.
We obtain
\be 
B^{-1}=M^3\Bigl((-4k+r_cp^2)K_2(p/k)-2p \, K'_2(p/k)\Bigr).
\label{bb} 
\ee
One may use the identity $K'_2(z)+2K_2(z)/z=-K_1(z)$ to rewrite
\be
B^{-1}=M^3 p \Bigl(2K_1(p/k)+r_c p K_2(p/k)\Bigr).
\label{B}
\ee
As we are interested in correlations on the brane we need
\be
G(p,z=0)= {1\over{M^3 p}}\;{{K_2(p/k)}\over{2K_1(p/k)+r_c p K_2(p/k)}}.
\label{g0} 
\ee

We first reproduce the results of the RS case $r_c=0$. With $G(p,z)$
depending only on the magnitude $p$ of the momentum, the corresponding 
potential due to the unit mass at $(x=0,z=0)$
as viewed by an observer on the 3-brane, at a distance $r$ from the source, is
\be
V(r)={1\over{2\pi^2 r}}\int_0^\infty dp\, p \sin{pr}\, G(p,z=0)
\label{V(r)}.
\ee
For the case at hand this leads to
\be
V(r)=\frac{1}{4\pi^2}\frac{k}{M^3r}
\int_0^\infty d{\tilde p} \frac{K_2(\tilde p)}{K_1(\tilde p)} \sin({\tilde p}kr)
=\frac{1}{4\pi}\frac{k}{M^3}\frac{1}{r} +\delta V(r),
\label{V} 
\ee
where 
\be 
\delta V(r)=\frac{1}{4\pi^2}\,\frac{k}{M^3r}
\int_0^\infty d{\tilde p} \frac{K_0(\tilde p)}{K_1(\tilde p)} \sin({\tilde p}kr).
\label{dV} 
\ee
In order to obtain the second equality above we used the identity 
$K_2(z)=2K_1(z)/z+K_0(z)$.

For ${\tilde p}\to\infty$ the ratio $K_0(\tilde p)/K_1(\tilde p)\to 1$ 
and the integral in (\ref{dV})
reduces to the ill-defined integral of $\sin{\tilde p}$ over the positive real axis.
In order to evaluate (\ref{dV}) we multiply the integrand by 
$e^{-q\tilde p}$, perform the integration and then take the limit $q\to 0$ 
\footnote{The integral depends on the regulator. The result remains the
same if instead of $e^{-q\tilde p}$ one
multiplies by $e^{-\alpha {\tilde p}^2}$ and then takes the limit $\alpha\to 0$,
while a simple ultraviolet cut-off $L$ on the ${\tilde p}$ leads
to an ambiguous answer.  
A correct regulator should reproduce
the well known answer $V(r) \sim 1/r^2$ for the case $G(p,0)\sim 1/p$,
corresponding to five-dimensional behavior.
The one we use satisfies this criterion. Equivalently, one could just use 
$\int_{-\infty}^{\infty} dx e^{i\alpha x}=2\pi\delta(\alpha)$ to 
define the umbiguous integrals mentioned above.}.

For $kr\gg 1$ the strongly oscillatory behavior of $\sin(\tilde p kr)$ results
in a negligible contribution to the integral from large $\tilde p$. 
This means that we can employ the expansion of the Bessel functions for 
small $\tilde p$: $K_0(\tilde p)/K_1(\tilde p)=-\tilde p \log\tilde p$. 
We obtain \cite{rs}
\be
\delta V(r)\simeq\frac{1}{8\pi}\,\frac{1}{M^3k}\,\frac{1}{r^3}.
\label{dv1} \ee
Thus we reproduce the leading and subleading behavior of the potential
at long distances in the RS scenario. 
For $kr\ll 1$ the main contribution to the integral comes from large $\tilde p$, for which
$K_0(\tilde p)/K_1(\tilde p)=1$. We find 
\be
\delta V(r)\simeq \frac{1}{4\pi^2}\,\frac{1}{M^3}\,\frac{1}{r^2}.
\label{dv2} \ee
Thus, at short distances gravity is $five$-$dimensional$  \cite{Kakushadze:2001rz}.
Moreover we see that the behavior of the potential at small distances 
is different from the one implied by the subleading term for large $kr$.

We next study the general case $r_c\not= 0$. The Fourier transform is 
difficult to perform explicitly, but we can deduce the behavior of the potential by
studying the behavior of the propagator as a function of $p$. 
We find that 
\be
{\rm for}~~~p \ll k, ~~~~~~~~~~~~~~~~
G(p,z=0)\simeq \frac{1}{M^3\left(r_c+\frac{1}{k}\right)p^2},
\label{g1} 
\ee
\be
{\rm for}~~~p \gg k, ~~~~~~~~~~~~~~~~
G(p,z=0)\simeq \frac{1}{M^3\left(r_c p^2+2p \right)}.
\label{g2} 
\ee
Clearly, for $r_c\not=0$ significant modifications of
the gravitational potential are possible. 

We will distinguish two separate cases: 

(a) $kr_c \gg 1$: Both for 
$p\ll k$ and $p\gg k$ we have $G^{-1} \simeq M^3r_cp^2$.
Thus we expect four-dimensional behavior $\sim 1/r$ for the potential at 
all distances on the brane, with an effective 
Planck constant $M^2_{Pl} \simeq M^3 r_c$. 

The leading corrections to $V(r)$ can also be evaluated 
by employing the full propagator (\ref{g0}). We find 
\begin{eqnarray}
{\rm for}~~
r \gg 1/k,~~~~~\delta V(r) &=& \frac{1}{8\pi} \frac{1}{M^3k} 
\frac{1}{(r_ck+1)^2} \frac{1}{r^3} 
\nonumber \\
&\simeq& 
\frac{1}{8\pi} \frac{1}{M^2_{Pl}} 
\frac{1}{r_ck^3} \frac{1}{r^3}
\simeq 
\frac{1}{8\pi} \frac{1}{M^4_{Pl}} 
\left(\frac{M}{k}\right)^3 \frac{1}{r^3},
\label{corre1} \\
{\rm for}~~
r \ll 1/k,~~~~~\delta V(r) &=& \frac{1}{\pi^2}
\frac{1}{M^3r^2_c} \log (kr) 
\nonumber \\
&\simeq& \frac{1}{\pi^2}
\frac{1}{\mpl^2r_c} \log (kr)
\simeq \frac{1}{\pi^2}
\frac{M^3}{\mpl^4} \log (kr).
\label{corre2}
\end{eqnarray}

(b) $kr_c \ll 1$: For $p\ll k$, $G^{-1} \simeq M^3p^2/k$. We expect that 
at large distances $r \gg 1/k$ the potential displays four-dimensional behavior
with $\tilde \mpl^2\simeq M^3/k$, as in the standard RS scenario. 
For $k \ll p \ll 1/r_c$, we have $G^{-1} \simeq 2 M^3 p$. 
Thus, for distances
$r_c \ll r \ll 1/k$ we expect
five-dimensional behavior $\sim 1/r^2$ for the potential. 
This is in agreement with
the direct evaluation of the potential for $r_c=0$. Finally, for $p \gg 1/r_c$,
$G^{-1} \simeq M^3 r_c p^2$. At short distances
$r \ll r_c$ the behavior becomes again four-dimensional $\sim 1/r$, with
$M^2_{Pl} \simeq M^3 r_c$. 

To summarize, the four-dimensional Einstein term 
induced quantum mechanically on the 3-brane affects considerably 
the gravitational interactions on the brane. Specifically, the gravitational 
potential on the brane exhibits the four-dimensional behavior $V(r)\sim 1/r$, except in the
intermediate region $r_c\ll r\ll 1/k$, in which it is effectively
five-dimensional, given by $V(r)\sim 1/r^2$. Furthermore, for $kr_c\ll 1$ the strength 
of the gravitational interaction, i.e. the value of the effective $M_{Pl}$, 
depends on the distance between the interacting masses. It is stronger for short distances,
the ratio of its value for $r\ll r_c$ to the one for large $r\gg 1/k$ being equal to $kr_c$.

\section{Kaluza-Klein modes and their effects}
\setcounter{equation}{0}

The physical picture of the previous section can also be 
confirmed by considering 
the massive gravitons, i.e. the Kaluza-Klein modes of the effective 
compactification induced by the warped geometry.
We ignore as before the tensor structure of the metric and denote by $\Phi(x^\alpha,z)$ 
the small fluctuation field around the background (\ref{rs1}). 
Its equation of motion at the linearized level is
\be
M^3\left[{1\over\sqrt{-g}}\partial_\mu\left(\sqrt{-g}g^{\mu\nu}\partial_\nu\right)
+\delta(z){{r_c}\over\sqrt{-{\hat g}}}\partial_\alpha
\left(\sqrt{-{\hat g}}{\hat g}^{\alpha\beta}
\partial_\beta\right)\right]\Phi(x^\alpha,z)=0.
\label{Phieqn}
\ee
As suggested by the symmetries of the background, we look for solutions in the form
$\Phi(x^\alpha,z)=\sum_n \phi_n(z)\sigma_n(x^\alpha)$, where
the $\sigma_n(x^\alpha)$ satisfy the four-dimensional
Klein-Gordon equation $(\partial^\alpha
\partial_\alpha+m_n^2)\sigma_n=0$. Using this in (\ref{Phieqn}), one is led to the field
equation
\begin{equation}
\left( \partial_z^2+e^{-2A}m_n^2+4A'\partial_z
+r_c \delta(z)m_n^2 \right) \phi_n(z)=0
\label{31}
\end{equation}  
for the mode function $\phi_n(z)$.
 
The zero mode, the solution corresponding to $m^2=0$, is not affected by
the presence of the term proportional to $r_c$ and consequently is identical
to the one in reference \cite{rs}.
The effective four-dimensional squared Planck constant is determined by
taking the low-energy limit of (\ref{00}). The term $\sim M^3 R$ 
gives $(M^3/k) \hat{R}$ in the low-energy effective action \cite{rs}. 
This contribution combined with the term 
$M^3 r_c \hat{R}$ results in a low-energy theory with an effective
squared Planck constant $M^3(r_c+1/k)$, in agreement with 
eq. (\ref{g1}).

The KK modes, 
analogous to those of \cite{rs}, 
are $\psi_n=\exp(3A/2)\,\phi_n$.
For $A(z)=-k|z|$ eq. (\ref{31}) gives 
(for simplicity we omit the index $n$ from
$m_n$ and $\psi_n$)
\begin{equation}
\psi(z)=N(\mt) w^{1/2} \left[Y_2(\mt w) + F(\mt) \, J_2 (\mt w)\right],
\label{solkk}
\end{equation}
with $w\equiv\exp(k|z|)$, $\mt=m/k$. The constant $F(\mt)$ 
is fixed by the discontinuity 
in $\partial_z \phi(0)$ due to the presence of the $\delta$-function
\be
F(\mt)=-\frac{2Y_1(\mt)+\rct \mt Y_2(\mt)}{2J_1(\mt)+\rct \mt J_2(\mt)},
\label{ccc}
\ee
with $\rct=r_ck$, and $Y_n$, $J_n$ the Bessel functions in standard notation.

For $w \to \infty$ the KK modes become approximately plane waves 
\begin{eqnarray}
\lefteqn{
\psi(w)\simeq N(\mt) \sqrt{\frac{2}{\pi \mt}}
\left[
\sin \left( \mt w - \frac{5}{4}\pi \right)
+ F(\mt) \cos \left( \mt w - \frac{5}{4}\pi \right)
\right] }
\nonumber \\
&\simeq&
N(\mt) \sqrt{\frac{2\left(1+F^2(\mt)\right)}{\pi \mt}}
\sin \left( \mt w - \frac{5}{4}\pi  + \beta(\mt) \right),
\label{plane} 
\end{eqnarray}
with $\beta = \arctan F$.
As a result, 
for a non-compact fifth dimension the KK modes have a continuous spectrum and
their normalization is
approximately that of plane waves
\be
N(\mt) \sim \sqrt{\frac{\mt}{1+F^2(\mt)}},
\label{norm} 
\ee
where we have neglected factors of order 1. 
The strength of the interaction of the KK graviton modes with the other fields
on the brane is determined by 
the square of their wavefunction at the position $z=0$ of the brane. We find
\be
\psi(z=0) \sim \sqrt{\frac{\mt}{1+F^2(\mt)}}
\left[Y_2(\mt)+F(\mt) \, J_2(\mt)\right].
\label{supp} 
\ee

A careful examination of the low energy effective action reveals the
presence of an additional suppression factor. The second term in 
the action (\ref{00}) results in a non-canonical kinetic term for
the fields $\sigma_n(x^\alpha)$. In order to render this term canonical
we must absorb a factor $(1+\rct |\psi(0)|^2)^{1/2}$ into the redefinition
of the fields. The proof is completely analogous to the derivation of 
eq. (7.16) of \cite{dgkn}. 
This results in a suppression of all interactions with external
sources by the same factor. However, this correction is negligible in
all the cases we study below. 

After the KK kinetic terms have been properly normalized,
the coupling of the KK modes to matter on the brane is 
given by $\sqrt{k/M^3}$. This coupling is squared in the 
calculation of quantities such as KK mode production rates etc.
It is also accompanied by the integration over all KK states with a
plane-wave measure $dm/k$. As a result, the combination $dm/M^3$ 
appears in all the estimates of KK mode production in the following.

For $\mt \ll 1$ eq. (\ref{supp}) gives
\be
\psi(z=0) \sim \frac{\sqrt{\mt}}{1+\rct}.
\label{rang1} \ee
We thus recover the suppression $\sim \sqrt{m/k}$ of the 
standard Randall-Sundrum model, which is further enhanced for
large $kr_c$.

For $\mt \gg 1$, on the other hand, we find
\be
\psi(z=0) \sim 
\left[ 1+\left( \frac{\rct \mt}{2} \right)^2 \right]^{-1/2}.
\label{rang2} 
\ee
For $m\gg 1/r_c$ there is a significant suppression factor
$\sim 1/(mr_c)$, while for $m \ll 1/r_c$
the wavefunction on the brane is unsupressed of order 1.

As we will show next, these results are consistent with the findings 
of the previous section and, in addition, crucial to clarify the origin 
of the behavior of the effective gravitational interaction on the brane.
We will again separate the two different cases: 
(a) $kr_c\gg 1$, where gravity was found to be four-dimensional at all distances;
(b) $kr_c\ll 1$, where gravity again behaves as in four dimensions,
except in the intermediate range $(k,1/r_c)$ where it is five-dimensional.

(a) $kr_c\gg 1$: For 
$m\lta k$ the wavefunction of the KK modes on the brane is
$\sim \sqrt{m/(r_c^2 k^3)}$, while for $m\gta k$ it is 
$\sim 1/(mr_c)$. 
Thus, the gravitational
potential is dominated by the exchange of the zero mode and falls off 
$\sim 1/r$ for all $r$. 
The effective squared Planck constant is $M^3(r_c + 1/k)\simeq M^3r_c$.

(b) $kr_c\ll 1$: 
The wavefunction of modes with $m\lta k$ is $\sim \sqrt{m/k}$, 
that of modes with $k \lta m \lta 1/r_c$ is of order 1, while
that of modes with $m \gta 1/r_c$ is $\sim 1/(mr_c)$.\\
i) For $r\gta 1/k$ 
the corrections to the four-dimensional 
potential are dominated by
modes with $m\lta k$ because the contribution of modes with $m\gta k$ 
is exponentially suppressed.
The contribution of massive modes 
is negligible relative to that of the
zero mode. Thus we expect a 
fall-off $\sim 1/r$
with a squared Planck constant $M^3(r_c + 1/k)\simeq M^3/k$.\\
ii) For $r_c \lta r \lta 1/k$ only the modes with $m \lta 1/r_c$ contribute 
significantly.
Those with 
$k \lta m \lta 1/r_c$ 
generate a term in the potential
\be
\delta V(r) \sim \frac{1}{M^3} \int_k^{1/r_c} dm \frac{e^{-mr}}{r} \psi^2(0)
\sim \frac{1}{M^3} \int_k^{1/r_c} dm \frac{e^{-mr}}{r}
\simeq \frac{1}{M^3r^2}. 
\label{corr1} \ee
This contribution is much larger than those from
the modes with $m \lta k$ and the zero mode.
For example the modes with $m \lta k$ give
\be
\delta V(r) \sim \frac{1}{M^3} \int_0^{k} dm \frac{e^{-mr}}{r} \frac{m}{k}
\simeq \frac{k}{M^3r}.
\label{corr2} \ee
Thus, for distances
$r_c \lta r \lta 1/k$ we expect
five-dimensional behavior $\sim 1/r^2$ for the potential. \\
iii) Finally, at short distances
$r \lta r_c$ the modes with $m \gta 1/r_c$ give a contribution
\be
\delta V_1(r) \sim \frac{1}{M^3} \int_{1/r_c}^\infty dm \frac{e^{-mr}}{r}
\frac{1}{m^2 r^2_c}
\simeq \frac{1}{M^3r_cr}. 
\label{corr3} \ee
Those with $1/r_c \gta m \gta k$ give 
\be
\delta V_2(r) \sim \frac{1}{M^3} \int^{1/r_c}_k dm \frac{e^{-mr}}{r}
\simeq \frac{1}{M^3r_cr}. 
\label{corr4} \ee
These dominate over the contribution of the zero mode, as well as
of the modes with $m \lta k$.
Thus, the potential obtains again the four-dimensional form $\sim 1/r$, with
a squared Planck constant $M^3 r_c$. It is remarkable that this behavior is not
due to the zero mode, as on might have guessed, but instead it is attributed to
the exchange of massive modes with masses $m\gta k$.
Similar behavior was also observed in \cite{Dvali:2000rv,Karch:2001ct}.

\section{Cosmology}
\setcounter{equation}{0}

The Friedmann equations for the RS cosmology, as modified by 
the induced Einstein term,  
were derived by Deffayet \cite{def} following the methods of \cite{betal}.
Here, we will quote the result and analyse it in 
a context not studied in the existing literature \cite{dvg}. 
The presence on the brane of matter/radiation and of a non-vanishing cosmological
constant, with total energy density $\rho_b$ and pressure $p_b$, results in a 
four-dimensional Friedmann-Robertson-Walker solution for the metric of the form
\be
ds^2=e^{2A(z)}\left[-dt^2+a^2(t)\left({{dr^2}\over{1-\lambda r^2}}+r^2d\theta^2+
r^2\sin^2\theta d\phi^2\right)\right]+dz^2.
\label{frw}
\ee
The scale factor $a(t)$, the Hubble constant $H=\dot a/a$ as viewed by an 
observer on the brane, and the spatial curvature $\lambda=\pm 1, 0$ 
are related by the Friedmann
equation
\be
{r_c^2\over 2}\left(H^2+{{\lambda}\over a^2}\right)
=1+{r_c\over 12M^3}\rho_b+\epsilon
\sqrt{1+{r_c\over 6M^3}\rho_b-{r_c^2\over 12 M^3}\Lambda+{r_c^2C\over a^4}}.
\label{friedman1}
\ee
$C$ is an  integration constant \cite{betal}. 
If non-zero, 
it generates a ``mirage'' radiation density on the brane \cite{mirage}.
In this work we set $C=0$.

The parameter 
$\epsilon=\pm 1$ defines two possible branches
in the solution. Its value is determined by the sign
of $dA/dz$ at $z=0$ \cite{def}, 
which must be negative for the graviton zero mode
to be localized on the brane. 
Thus, cosmology on a brane with four-dimensional gravity at large distances
requires $\epsilon=-1$.
As this is the scenario of interest to
us, we set $\epsilon=-1$ in the following.
The case $\epsilon=1$ was discussed in \cite{dvg} where it was pointed out that it
generates a late time cosmological constant.

For simplicity we will assume, for the time being, that there in no significant 
flow of energy out of the brane through the decay of brane matter into KK modes of
the graviton \footnote{The validity of this assumption is not guaranteed for 
large energy densities when $k\ll 1/r_c$, as
the KK modes with $k\lta m \lta 1/r_c$ are unsuppressed on the brane.
We will come back to this point in the next section, where we 
will estimate the energy loss.}. 
Under this assumption, the energy density $\rho_b$ on the brane
satisfies the conservation equation
\be
\dot \rho_b=-3H(\rho_b+p_b).
\ee

We are studying here the effects induced by the presence 
of brane matter and radiation on the 
RS ``vacuum" background (\ref{rs1}) with a given value of $k$,
i.e. with the parameters $\Lambda$ and $V_b$ of the theory satisfying 
$V_b=-\Lambda/k=12M^3k$. As a consistency check, it is straightforward to verify
that (\ref{friedman1}) is satisfied by (\ref{rs1}) for 
$\rho_b=V_b=-p_b$, $\epsilon=-1$, and $C=\lambda=0$. 
Separating the energy density as
$\rho_b=V_b+\rho$ and using the above values for $\Lambda$ and $V_b$, (\ref{friedman1})
takes the form
\be
{r_c^2\over 2}\left(H^2+{{\lambda}\over a^2}\right)
=1+kr_c+{r_c\over 12M^3}\rho-
\sqrt{(1+kr_c)^2+{r_c\over 6M^3}\rho}.
\label{friedman2}
\ee
This equation has a smooth limit as $r_c\to 0$ 
and gives the cosmological evolution
of the RS universe
\be
H^2={k\rho\over 6M^3}+{1\over 4}\left({\rho\over 6M^3}\right)^2
-{\lambda\over a^2}.
\ee

In analysing the physical content of (\ref{friedman2}) we will distinguish
two cases: 

(a) The case $kr_c\gg 1$. For the 
gravitational potential
this corresponds 
to four-dimensional
behavior on the brane at all scales with $\mpl^2=M^3r_c$.
We define the normalized density $\tr\equiv {r_c\rho/ M^3}$. We have
\\
(i)  for $\tr\gg (kr_c)^2$ or $\tr\simeq (kr_c)^2$ or $\tr\ll (kr_c)^2$:
\be
H^2\simeq {\rho\over 6\mpl^2}-{\lambda\over a^2},
\ee
Thus, we obtain at all times the 
standard Friedmann equation.

(b) The case $kr_c\ll 1$.
The gravitational potential on the brane is four-dimensional 
at energies $E\ll k$ with $\tilde \mpl^2=M^3/k$ (RS regime), 
and for $E\gg 1/r_c$ with $\mpl^2=M^3r_c$ (Induced Gravity (IG) regime).
At intermediate energies $k\ll E\ll 1/r_c$ gravity is five-dimensional 
($5d$ regime).
We obtain
\\
(i) $\tr\gg 1$ (or $\rho/M^3 \gg 1/r_c$):
\be
H^2\simeq {\rho\over 6\mpl^2}-{\lambda\over a^2},
\ee
corresponding to the IG regime.
\\
(ii) $\tr \ll 1$:
\be
H^2\simeq {\rho\over 6\tilde \mpl^2}
+{1 \over 4} \left( {\rho \over 6 \tilde \mpl^2 k} \right)^2 
-{\lambda\over a^2}.
\label{5d}
\ee
We recover the cosmology of the RS universe.
The range $\rho/M^3 \ll k$
corresponds to the RS regime
\be
H^2\simeq {\rho\over 6\tilde \mpl^2}
-{\lambda\over a^2},
\ee
while for the range  
$k \ll \rho/M^3 \ll 1/r_c$ the term $\sim \rho^2$ in eq. (\ref{5d}) 
is important ($5d$ regime).

Thus, we can confirm that 
the rough cosmological evolution is analogous to the static 
behavior of gravity on the brane. The various regimes can
be discussed in terms of an energy scale
set by $1/r$ or $\rho/M^3$, for the static potential or
cosmology respectively.
However, we have neglected a potentially  important factor: the 
energy that is radiated off the brane 
into the bulk in the form of KK gravitons.
We will study this in more detail in the next section.

\section{Emission of Kaluza-Klein modes and experimental constraints}
\setcounter{equation}{0}

Matter on the brane can loose energy by emitting KK gravitons.
Here we will estimate the efficiency of such processes in various contexts.
 
We will again separate the two different cases: 

(a) $kr_c \gg 1$. This case is not very constrained by experiment.
All KK modes are significantly 
suppressed and do not affect standard processes. For example,
the rate of emission of KK modes from a star can be estimated
as 
\be
\Gamma(T) \propto \frac{1}{M^3} \int_0^T dm\, \psi(0)^2 
\sim \frac{1}{M^3} \int_0^T dm \frac{m}{k} \frac{1}{r_c^2k^2}
\sim \frac{1}{\mpl^2} \frac{T^2}{k^2} \frac{1}{r_ck}
\label{rate1}
\ee
for $T <k$. This is much smaller than the rate of production 
of zero-mode gravitons $\Gamma_0(T) \propto 1/\mpl^2$, and is negligible.
For $T > k$ the largest contribution to the rate is 
\be
\Gamma(T) \propto 
\frac{1}{M^3} \int_k^T dm \frac{1}{m^2r_c^2} 
\sim \frac{1}{\mpl^2} \frac{1}{r_ck},
\label{rate2}
\ee
and is negligible again. Thus there are no severe constraints on
the parameters apart from the requirement to reproduce the
value of the Planck constant $\mpl^2 =M^3 r_c$.

The most interesting property of the scenario concerns the 
corrections to the gravitational potential. In particular, the form
of the corrections changes at a characteristic distance $\sim 1/k$,
according to eqs. (\ref{corre1}),(\ref{corre2}). Since $k$ is largely 
unconstrained by other considerations it is essentially
a free parameter.
The logarithmic variation at small distances is 
probably too slow to be detected. The $1/r^3$ correction to Newton's law 
at large distances is constrained by experiment \cite{powerlaw,sav2}. 
If the potential is parametrized as
\be
V(r)= C N_1 N_2 \frac{(10^{-15}~{\rm m})^2}{r^3},
\label{exp} \ee
with $N_1$, $N_2$ the number of nucleons of the two interacting objects,
the strongest bound is $C \leq 7 \times 10^{-7}$. Comparing with 
eq. (\ref{corre1}) we obtain $k/M \gta 10^{-20}$. 

Since the four-dimensional Planck scale is related to $r_c$ for $r_c \gg 1/k$
there are enough free parameters to make the values of $\Lambda$, $V_b$
natural.
Choosing $k=M=M_{\rm SUSY}$ indeed accomplishes this.
$r_c$ is determined from $M_{Pl}^2=M^3r_c$.
However, it should be remembered that, in a given theory, 
$r_c$ is determined by the loop corrections and is
not a free parameter. 

The cosmological evolution in this scenario is standard for 
all densities. However, there is a small amount of energy loss to the
bulk. The change in energy density 
per unit time is equal to the 
rate of energy loss to KK gravitons per unit time and volume.
For a process $a+b \to c + KK$ it is given by 
\be
\left( \frac{d\rho}{dt} \right)_{\rm lost} 
= - \left\langle n_a \, n_b\, \sigma_{a+b \to c + KK}\, v \,E_{KK} 
\right\rangle,
\label{energyloss1} \ee
where the brackets indicate thermal averaging.
For a radiation-dominated brane 
we can take approximately $n_a,n_b\sim T^3$, $E_{KK}\sim T$,
and estimate 
\be
\left\langle \sigma_{a+b \to c + KK}\, v \right\rangle
\sim \frac{1}{M^3} \int_k^T dm \frac{1}{m^2r_c^2} 
\sim \frac{1}{\mpl^2} \frac{1}{r_ck},
\label{sigmavt} \ee
in agreement with eq. (\ref{rate2}).
This leads to 
\be
\left( \frac{d\rho}{dt} \right)_{\rm lost} 
\sim - \frac{T^7}{\mpl^2} \frac{1}{r_c k}
\sim - \frac{\rho^{7/4}}{\mpl^2} \frac{1}{r_c k}.
\label{enloss} \ee
We conclude that the energy loss is negligible because
$(d\rho/dt)_{\rm expansion}/(d\rho/dt)_{\rm lost} \sim 
(\mpl/\rho^{1/4})r_ck\gg 1$. Thus, 
we obtain a standard Friedmann cosmological expansion with essentially 
no energy loss.

(b) For $kr_c \ll 1$ the deviations from the standard RS physics appear at
energy scales much larger than $k$. For the gravitational potential,
we expect a transition from the four-dimensional form $\sim 1/r$ 
to the five-dimensional one $\sim 1/r^2$ at distances $r\lta  1/k$. 
The experimental constraints require $k \gta (10~\mu{\rm m})^{-1} \simeq
10^{-11}$ GeV, while the value of $M$ is
fixed by the relation $\tilde \mpl^2 = M^3/k$ to be
$M\gta  10^9$ GeV.

The emission of KK modes with masses 
$1/r_c \gta m \gta k$ is unsuppressed on the brane. Their contribution to
various processes, such as star cooling or high-energy experiments,
is analogous to those in standard torroidal compactifications,  
\cite{sav2,sav1} with one extra dimension and a compactification radius $\sim 1/k$. 
The strongest constraints arise from star cooling through the emission of
KK modes. 
For $1/r_c>T>k$ the largest contribution is
\be
\Gamma(T) 
\propto \frac{1}{M^3} \int_0^T dm\, \psi(0)^2
\sim \frac{1}{M^3} \int_k^T dm 
\sim \frac{1}{\tilde \mpl^2} \frac{T}{k}.
\label{rate00}
\ee
For the case of a supernova with $T \sim 30$ MeV, we can require that
this rate be smaller than the equivalent axion production rate 
$\Gamma_a \propto 1/f_a^2$, for which the constraint is
$f_a \gta 10^9$ GeV \cite{sav2}. This leads to $k\gta 10^{-21}$ GeV, a much
weaker bound than the one imposed by deviations from Newtonian gravity.
In a more careful treatment one should take into account the details of
the various processes of KK graviton production, such as nucleon-nucleon
Brehmstrahlung, gravi-Compton scattering etc. However, the basic conclusion
that the resulting bound is much weaker than the one imposed by 
deviations from Newtonian gravity is expected to be unaffected.
The situation  with $T> 1/r_c$ leads to more relaxed constraints because
of the suppression of KK modes with $m \gta 1/r_c$ on the brane.

For $k\gta 10^{-11}$ GeV, $M\gta 10^{9}$ GeV
we can evaluate the vacuum energy in the bulk 
$|\Lambda| \gta (10$ GeV)$^5$ and the brane 
$V_b \gta (10 $ TeV)$^4$.
The length $r_c$ is not constrained and determines the weakness 
of gravity at short distances
$\tilde \mpl/\mpl=\sqrt{kr_c}\ll 1$.

Note that the lowest values of $V_b$ are compatible with 
the possibility that
it arises through supersymmetry breaking
on the brane with $M_{\rm SUSY}$ of the order of 10 TeV.
Moreover, it is expected that it will introduce a bulk supersymmetry breaking
scale 
of the order of $M_{\rm SUSY}^2/M\sim 0.1$ GeV.
This scale is about 100 times smaller than the one needed to equilibrate and
flatten the brane.

The cosmology of this scenario has several novel features, as was indicated
in the previous section. 
For densities $\rho \lta M^3 k$ one expects the standard cosmological
evolution with $H^2 \sim \rho/{\tilde \mpl^2}$ (RS regime). 
For $k$, $M$ near the lower 
bound set by observations $k \sim 10^{-11}$, $M\sim 10^9$ GeV, this
regime extends up to densities $\rho \sim (10\,{\rm TeV})^4$.
However, for 
$M^3 k \lta \rho \lta M^3/r_c $ the Hubble parameter behaves
$H^2 \sim \rho^2/M^6$ ($5d$ regime), while for $\rho \gta M^3/r_c$ we have
$H^2 \sim \rho/\mpl^2$ (IG regime). 

As shown in the previous section, unsuppressed emission of single KK gravitons 
can take place for the mass range $1/r_c> m> k$.
For a brane with energy density $\rho \gta k^4$ it is 
possible to produce such unsuppressed KK gravitons 
that escape into the bulk.
We concentrate on the case of a radiation-dominated brane-universe
($\rho\sim T^4$), which is the most relevant for the energy scales of 
interest. The scale 
$k^4$ is smaller than $M^3k$ because we assume 
$k\ll M$ (otherwise the 
whole energy regime above $k$ is strongly coupled).
We also assume $Mr_c \gg 1$ (otherwise
induced-gravity effects are masked by strong five-dimensional gravity).

\begin{figure}[htb]
\centering
\epsfxsize=5.7in
\epsfysize=1.5in
\epsffile{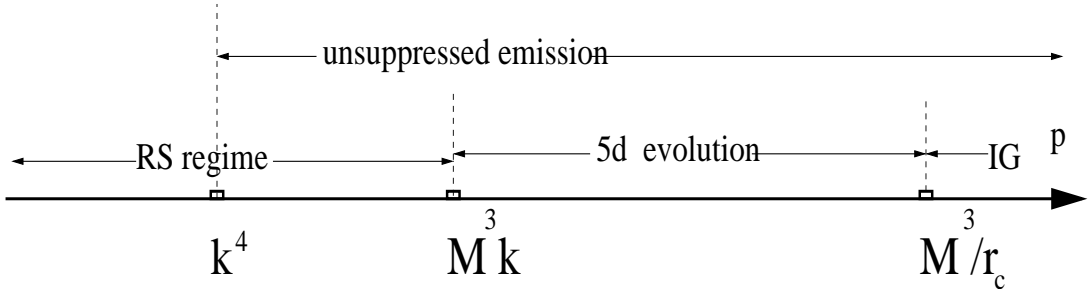}
\vspace{0.0cm}
\caption{Different regions in density, during the evolution of the universe. 
}
\label{sin1}
\end{figure}

The energy lost through emission of unsuppressed KK gravitons is given by 
eq. (\ref{energyloss1}). We estimate
\be
\left\langle \sigma_{a+b \to c + KK}\, v \right\rangle
\sim \frac{1}{M^3} \int_k^{\min(T,1/r_c)} dm = \frac{\min(T,1/r_c)}{M^3}.
\label{sigmavt2} \ee
For a given temperature $T$, the energy loss is maximized if $1/r_c > T$. 
We concentrate on this case in the following and find
\be
\left( \frac{d\rho}{dt} \right)_{\rm lost} 
\sim - \frac{T^8}{M^3} \sim - \frac{\rho^2}{M^3}.
\label{energyloss} \ee
The change in energy density because of the expansion is
\be
\left( \frac{d\rho}{dt} \right)_{\rm exp} 
= -4 H \rho.
\label{expansion} \ee

In the RS regime 
$(d\rho/dt)_{\rm exp}/(d\rho/dt)_{\rm lost} \sim (M^3k/\rho)^{1/2}\gg 1$.
Thus the energy loss during this period is negligible compared 
to the dilution due to the expansion.
In particular, for $M^3 k \sim (10\,{\rm TeV})^4$, the energy loss during 
nucleosynthesis is smaller by a factor $10^{14}$ than the rate of decrease 
of the 
energy density because of the expansion. Thus, during this period
the standard cosmological 
evolution  is not affected by energy loss.
A differed way of saying this is that graviton emission is frozen out
during the RS period.

The energy loss is substantial during the $5d$ regime, when
$(d\rho/dt)_{\rm exp} \sim -\rho^2/M^3$. 
Both $(d\rho/dt)_{\rm exp}$ and $(d\rho/dt)_{\rm lost}$ are of the same
order of magnitude and both lead to a decrease of the energy density 
for an expanding universe. 
The Friedmann equation (\ref{friedman1}) is not applicable any more.
The study of the cosmological evolution
requires the solution of Einstein's equations
in the presence of a significant off-diagonal term $T^5_0$ in the 
energy-momentum tensor. 
A first integral of these equations 
(leading to the generalization of (\ref{friedman1}))
is difficult to obtain. Work in this direction is under way.

In our discussion we assumed that $1/r_c>T$, so that the energy loss is 
maximized. (For this to hold during part of the $5d$ regime, 
we must have $r_c^{-4}> M^3 k$.)   
In the opposite case $1/r_c>T$, the energy loss is smaller by a 
factor of $r_c T$. Finally, it can be checked that the energy loss is
smaller than the energy dilution through expansion during the IG regime if
$1/r_c < M$.

\section{Conclusions}
\setcounter{equation}{0}

We have analysed some of the physics of the RS 3-brane universe embedded in five dimensions
in the presence of the four-dimensional Einstein term, induced by quantum loops
of particles confined on the brane. 
The scenario is characterized by the two length scales
$1/k$ and $r_c$, associated with the cosmological constants and the induced Einstein
term respectively. As we argued, the presence of the latter may affect 
considerably the short-distance structure of gravity on the brane.

Specifically, we showed that the four-dimensional $\hat R$ term in (\ref{001}) 
dominates the gravitational potential for all distances smaller than 
the characteristic scale $r_c$ associated with it, and results in the usual $\sim 1/r$
four-dimensional behavior. 
For $r\gg r_c$ its role is subdominant compared to the five-dimensional
Einstein term. Thus, it modifies the short distance 
behavior of the gravitational potential of the RS scenario from five- to 
four-dimensional, while it leaves
its large distance four-dimensional form unaffected.

Our conclusions for the gravitational interaction on the brane, 
for the two generic cases that parallel the 
situation with toroidal compactification 
plus induced four-dimensional gravity \cite{dgkn}, 
are summarized as follows:
\begin{itemize}
\item[(a)] 
$kr_c\gg 1$. Gravity on the 3-brane is four-dimensional at all distances. 
\item[(b)] 
$kr_c\ll 1$. Gravity on the 3-brane is four-dimensional at long and short distances, 
except for the intermediate range $(k,1/r_c)$ where it exhibits five-dimensional behavior.
\end{itemize}

We have analysed the spectrum and couplings of the KK modes of the graviton. 
We find that in the regimes where induced gravity is dominant there is a 
further supression of their couplings.
The couplings are not supressed only in the mass range that corresponds
to gravity having a five-dimensional behavior.

We have further analysed the cosmological evolution equations in the standard branch 
in the presence of bulk and brane cosmological constants. 
We show that in case (a) the cosmological evolution is very 
similar to the standard four-dimensional Friedmann evolution.
In case (b) there is an intermediate period where the 
cosmological evolution is 
five-dimensional \cite{betal}, namely $H^2\sim \rho^2$, 
while for $\rho\ll M^3k$ we recover the standard four-dimensional 
Friedmann equation $H^2\sim \rho$. There is also an earlier epoch 
during which the evolution 
is again four-dimensional but with a different Planck scale $M_{Pl}^2=M^3r_c$.  

An interesting effect that has to be added to the standard evolution equations
is the leakage of energy density from the brane to the bulk due to (massive) 
graviton emission. This has been estimated using the results of section 3.
It turns out that, 
for the low-density period in which the Friedmann equation is effectively 
four-dimensional ($H^2\sim \rho$),
the leakage contribution to the time derivative 
of the brane energy density is negligible compared to the 
``dilution" term $-3H(\rho +p)$.
Thus, normal four-dimensional intuition applies and no constraints 
from succesful late time events (nucleosynthesis, CMBR) are important.
During the $5d$ period on the other hand, the leakage term is as important 
as the dilution term.

With respect to the effect of phenomenological 
constraints on the parameters of the model,
the situation can be summarized as follows:

Case (a) is not severely constrained by experiment except for the requirement
to reproduce the 
Planck scale $M_{Pl}^2=M^3r_c$.
Constraints on the subleading terms of the gravitational interaction impose
$k/M\gta 10^{-20}$.
Moreover, there is enough free parameter space to make the values of bulk 
and brane cosmological constants natural. 
Choosing $k=M=M_{\rm SUSY}$ indeed accomplishes this.
However, it should be remembered that in a given theory $r_c$ is
not a free parameter, but is determined by the loop corrections.

Case (b) is more tightly constrained. From experimental constraints, 
the energy scale must satisfy $k\gta 10^{-11}$ GeV corresponding to
distances shorter than 10 nm.
Thus, at large distances the strength of gravity is determined 
by $\tilde M_{Pl}=\sqrt{M^3/k}\simeq 10^{19}$
GeV, implying $M\gta  10^9$ GeV.
We can evaluate the vacuum energy in the bulk 
$\Lambda\gta (10^{2/5}$ GeV)$^5$ and the brane 
$V_b\gta (10 $ TeV)$^4$.
The length scale 
$r_c$ is not constrained and determines the strength of gravity 
at short distances
$\tilde M_{Pl}/M_{Pl}=\sqrt{kr_c}\ll 1$.
Note that gravity becomes stronger at distances beyond $1/r_c$.
Moreover, the cosmological evolution is indistiguishable from the standard one 
except for very early times.

The lowest values of $V_b$ are compatible with the possibility that it
arises through supersymmetry breaking
on the brane with $M_{\rm SUSY}$ of the order of 10 TeV.
Moreover, it is expected that it will introduce a bulk supersymmetry breaking
scale of the order of $M_{\rm SUSY}^2/M\sim 0.1$ GeV.
This scale is about 100 times smaller than the one needed to equilibrate and
flatten the brane.

\vskip 1.5cm
\centerline{\bf\large Acknowledgments}
\vskip .5cm

We would like to thank G. Kofinas for useful discussions.
The work of N. Tetradis was  partially supported through a RTN contract
HPRN--CT--2000--00148 of the European Union.
The work of E. Kiritsis and T. Tomaras was partially supported by 
RTN contracts HPRN--CT--2000--00122 and --00131.
We acknowledge also partial support from INTAS grant N 99 1 590.

\end{document}